\RequirePackage{lineno}
\documentclass[3p,times]{elsarticle}

\usepackage{ecrc}
\usepackage{upgreek}

\volume{00}

\firstpage{1}

\journalname{Nuclear Physics A}

\runauth{}

\jid{nupha}

\usepackage{amssymb}

\usepackage[figuresright]{rotating}

\begin{document}

\begin{frontmatter}



\dochead{}

\title{J/$\psi$ elliptic flow measurement in Pb-Pb collisions at $\sqrt{s_{\rm NN}}$ = 2.76 TeV at forward rapidity with the ALICE experiment}


\author{L. Massacrier for the ALICE Collaboration}

\address{Universit\'e de Nantes, Ecole des Mines and CNRS/IN2P3, Nantes, France \\
\small{Contact : Laure.MASSACRIER@subatech.in2p3.fr}}

\begin{abstract}
J/$\psi$ suppression induced by color screening of its constituent quarks was proposed 26 years ago as a signature of the formation of a quark gluon plasma in heavy-ion collisions. Recent results from ALICE in Pb-Pb collisions exhibit a smaller suppression with respect to previous measurements at the SPS and RHIC. The study of azimuthal anisotropy in particle production gives information on the collective hydrodynamic expansion at the early stage of the fireball, where the matter created in high-energy nuclear collisions is expected to be in a deconfined state. In particular, J/$\psi$ elliptic flow $v_{2}$ is important to test the degree of thermalization of heavy quarks. Together with the production yields, the elliptic flow is a powerful observable to address the question of suppression and regeneration of J/$\psi$ in QGP.
We present the first inclusive J/$\psi$ elliptic flow measurement performed with the muon spectrometer of ALICE, in Pb-Pb collisions, at forward rapidity. Integrated and $p_{\rm T}$-differential $v_{2}$ results are presented and a comparison with recent STAR results and with a parton transport model is also performed.
\end{abstract}


\end{frontmatter}


\section{Introduction}

Heavy quarks produced in ultra-relativistic heavy-ion collisions are well known probes of the Quark Gluon Plasma, a state of matter created at high energy densities and predicted by Quantum Chromodynamics. Indeed, sequential quarkonium states suppression induced by color screening gives information on the initial temperature of the system \cite{MATS}. ALICE published recently the nuclear modification factor $R_{AA}$ of inclusive J/$\psi$ down to zero $p_{\rm T}$ at forward rapidity in Pb-Pb collisions at $\sqrt{s_{\rm NN}}$ = 2.76 TeV \cite{ALICEjpsi}. J/$\psi$ are found to be less suppressed in central collisions with respect to SPS \cite{SPS} and RHIC \cite{RHIC1,RHIC2} measurements. Statistical hadronization \cite{HAD} and transport models \cite{TM,TM2}, assuming J/$\psi$ regeneration from $c\bar{c}$ pairs, are able to reproduce ALICE data. The study of other observables, such as the azimuthal anisotropy of produced particles, brings additional information on how these particles are produced. In semi-central heavy-ion collisions, the distribution of the colliding matter is anisotropic in the plane perpendicular to the beam direction. If the matter is interacting, this spatial asymmetry is transferred to the momentum distribution of produced particles. The second moment of the final state hadron angular distribution is called the elliptic flow ($v_{2}$). The measurement of the J/$\psi$ elliptic flow is particularly interesting to learn about the interaction mechanisms within the QGP. 
Primordial J/$\psi$ produced by hard processes at the early stage of the collision are expected to be insensitive to collective phenomena. On the contrary, J/$\psi$ produced by recombination of charmed quark pairs in a deconfined phase should inherit the flow of charmed quarks, and consequently we expect to measure a non-zero elliptic flow. The STAR Collaboration reported the measurement of the J/$\psi$ elliptic flow in Au-Au collisions at $\sqrt{s_{\rm NN}}$ = 200 GeV performed at mid-rapidity \cite{STAR}. J/$\psi$ $v_{2}$ is found to be consistent with zero in all of the measured $p_{\rm T}$ range within uncertainties.
In this proceeding, we report on the first measurement of the inclusive J/$\psi$ $v_{2}$ performed with the ALICE detector, in Pb-Pb collisions at $\sqrt{s_{NN}}$ = 2.76 TeV in the $p_{\rm T}$ range 0-10 GeV/c and rapidity range $2.5 < y < 4$. The J/$\psi$ is studied via its dimuon decay channel. The evolution of the J/$\psi$ $v_{2}$ with centrality and $p_{\rm T}$ is presented.

\section{Experimental conditions}
The ALICE detector is decribed in \cite{ALICESetup}. The relevant detectors for this analysis are the muon spectrometer, the VZERO scintillators, the first two layers (SPD) of the inner tracking system (ITS), and the time projection chamber (TPC).
The muon forward spectrometer covers the pseudo-rapidity range $-4 < \eta < -2.5$ and consists of a front hadron absorber (10$\lambda_{I}$), five tracking stations composed of two planes of cathode pad chambers (CPC) each, a dipole magnet, an iron wall for filtering secondary and punch-through hadrons, and two triggering stations composed of two planes of resistive plate chambers (RPC) each. It allows the quarkonia measurement down to zero $p_{\rm T}$.
The VZERO detector is made of two scintillator arrays (32 scintillators each) located on both sides of the interaction point and covering pseudo-rapidity regions $-3.7 < \eta < -1.7$ (VZERO-C) and $2.8 < \eta < 5.1$ (VZERO-A). Each array is divided into 4 rings of 8 scintillators. The VZERO-C detector shares part of the muon spectrometer acceptance, therefore only the VZERO-A is used to determine the event plane to avoid auto-correlations. The VZERO detector is also used for the determination of the centrality of the collision. The SPD consists of two layers covering $|\eta| < 2.0$ and $|\eta| < 1.6$, respectively. It is used for triggering and to determine the vertex position of the Pb-Pb interaction.
The TPC is used to determine the event plane resolution of the VZERO-A. The minimum bias trigger condition in ALICE requires a signal in two readout chips in the outer layer of SPD, a signal in VZERO-A and a signal in VZERO-C. This trigger is downscaled to a rate of 10 Hz and the data sample collected amounts to an integrated luminosity of about 1 $\mu b^{-1}$. Events collected with this trigger are used for the flattening of the VZERO-A event plane angle distribution.
Complementary to the minimum bias trigger, a dimuon trigger is used. The dimuon trigger condition consists of the minimum bias trigger conditions in addition to the requirement of the detection of 2 tracks passing the low $p_{\rm T}$ cut given by the trigger on-line algorithm (50$\%$ efficiency for 1 GeV/c $p_{\rm T}$ tracks). In this analysis, the total integrated luminosity corresponds to 70 $\mu b^{-1}$.

\section{Analysis}
The J/$\psi$ elliptic flow was extracted using the event plane method. In this method, the azimuthal angle of the reaction plane is estimated from the observed event plane angle determined from the anisotropic flow itself. 
The event flow vector is defined as:
\begin{equation}
Q_{2,x} = \frac{\sum_{i=0}^{31}S_{i}cos(2\upphi_{i})}{\sum_{i=0}^{31}{S_{i}}} \hspace{1 cm} Q_{2,y} = \frac{\sum_{i=0}^{31}S_{i}sin(2\upphi_{i})}{\sum_{i=0}^{31}{S_{i}}}  
\end{equation}
where $\upphi_{i}$ is the mean azimuthal angle of each scintillator of VZERO-A and $S_{i}$ is a weight proportional to the flux of particles crossing those scintillators. 
The event plane angle for the $2^{nd}$ harmonic is then expressed as:
\begin{equation}
\Psi_{EP,2}=\frac{1}{2}\times\rm{tan}^{-1}\frac{Q_{2,y}}{Q_{2,x}}
\end{equation}
In order to correct for detector non-uniformities, a flattening procedure of the event plane azimuthal distribution is applied in two steps. The first step of the flattening procedure is described in \cite{FLAT1}. The second step uses a Fourier flattening technique \cite{FLAT2} and removes part of the residual effects due to the azimuthal segmentation of VZERO rings. A cut on the $z_{vertex}$ coordinate (direction along the beam axis) is also applied ($|z_{vertex}| <$ 10 cm) in order to improve the flattening performances. The following selection criteria are applied to the muons: both muon tracks reconstructed in the muon tracker should match a tracklet in the muon trigger stations and both muon tracks should exit the hadron absorber within the acceptance of the muon spectrometer. Opposite sign dimuons in the rapidity range $2.5 < y < 4$ are then selected. For each $p_{\rm T}$ and centrality bin the event sample is divided into 6 $\Delta\varphi$ bins, with $\Delta\varphi = \upphi_{dimuon} - \Psi_{EP,2}$ and the corresponding dimuon invariant mass spectra are fitted to extract the $J/\psi$ signal. In order to address the systematic uncertainties coming from the signal extraction, two different functions are used to fit the signal and the background (a Crystal Ball or a Crystal Ball with an additional tail on the higher mass region to describe the signal, a third order polynomial function or a variable width gaussian to describe the background). The width of the Crystal Ball in each $\Delta\varphi$ bin is fixed to the value obtained from a fit of the corresponding invariant mass distribution integrated over $\Delta\varphi$ with the Crystal Ball width as a free parameter. To evaluate the systematic uncertainties, the width of the Crystal Ball is fixed within the range of its error and the fitting range is also modified. The observed elliptic flow $v_{2}^{obs}$ is the Fourier coefficient extracted from the fit of the J/$\psi$ azimuthal angle distribution with the expression:
\begin{equation}
\label{eq}
\frac{dN^{J/\psi}}{d\Delta\varphi} = N_{0}[1 + 2 v_{2}^{obs} cos(2\Delta\varphi)]
\end{equation}

\begin{figure}[!htbp]
\begin{center}
\begin{minipage}[t]{.47\linewidth}
\hglue -0.5 true cm
 \includegraphics[width=\linewidth]{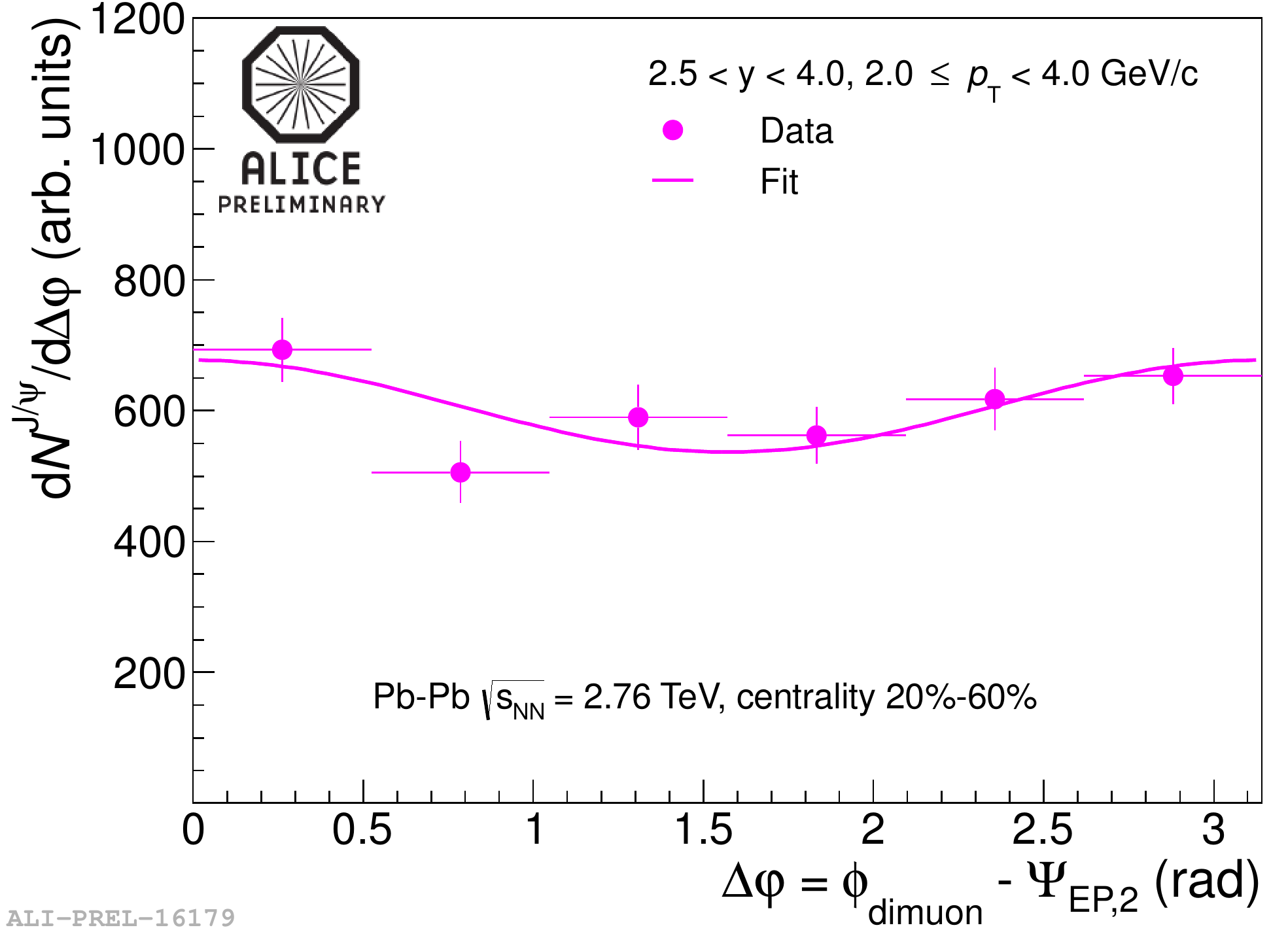}
\end{minipage}
\begin{minipage}[t]{.47\linewidth}
 \includegraphics[width=\linewidth]{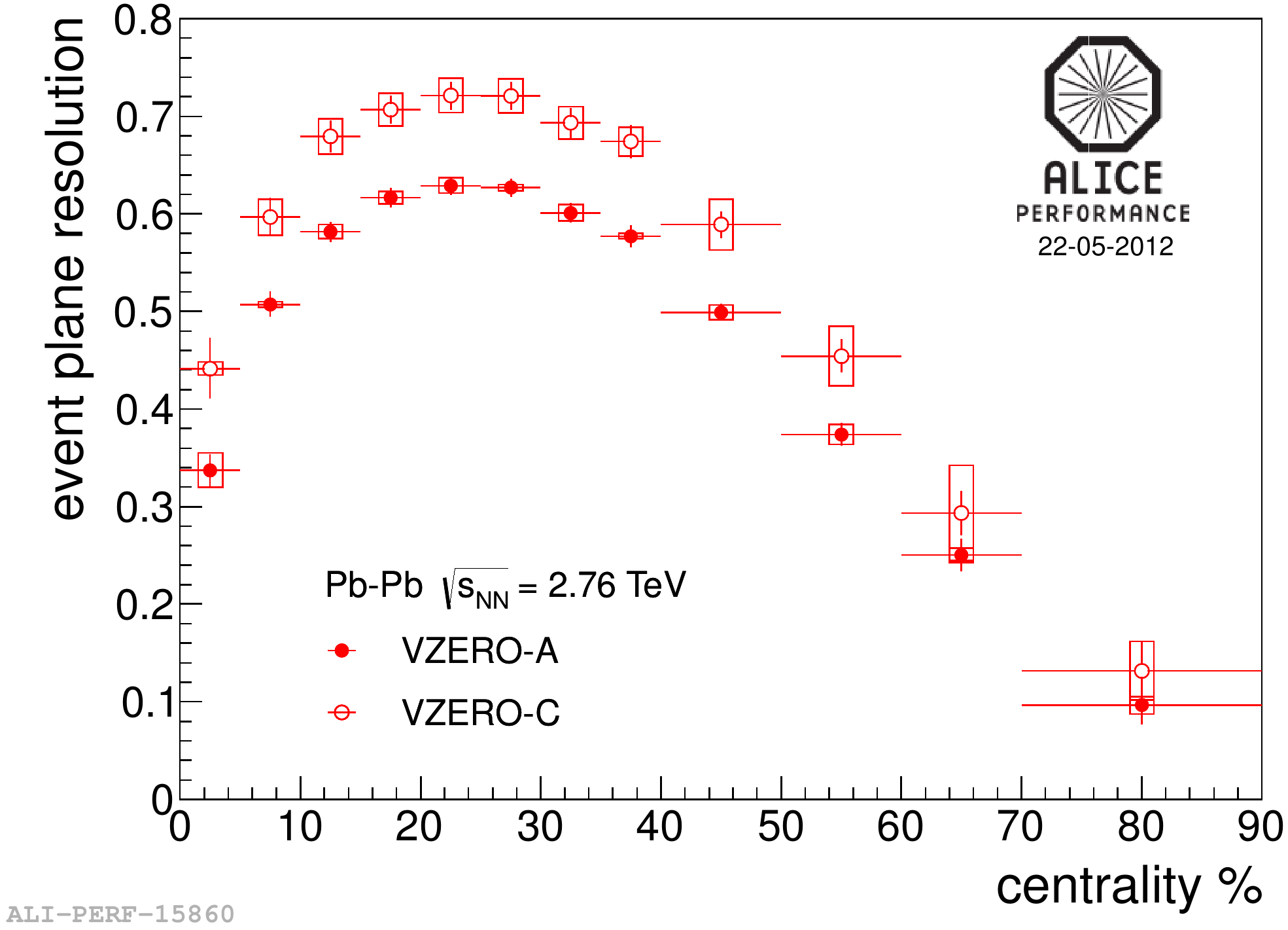}
\end{minipage}
\caption[fig1]{\label{fig1} \scriptsize{Left:  $dN/d\Delta\varphi$ distribution of J/$\psi$ for the centrality bin 20-60$\%$, 2 $\leq p_{\rm T} < $ 4 GeV/c and 2.5 $< y <$ 4. Data are fitted with the function in Eq. \ref{eq}}. Right: Event plane resolution evaluated with the three sub-events method as a function of centrality. VZERO-A event plane resolution is represented with full circles and VZERO-C resolution with empty circles. Boxes correpond to systematic uncertainties evaluated with two sets of sub-events. Crosses correspond to statistical uncertainties.}
\end{center}
\vglue -1 true cm
\end{figure}

Fig. \ref{fig1} left shows the dN/d$\Delta\varphi$ distribution of J/$\psi$ for the centrality range 20-60$\%$ and 2 $\leq p_{\rm T} <$ 4 GeV/c, fitted with Eq. \ref{eq}.  The observed $v_{2}^{obs}$ is corrected by the event plane resolution $R$ such as $v_{2} = v_{2}^{obs}/R$. The event plane resolution of the VZERO-A detector is determined using the three sub-events method \cite{POSK}. The event plane resolutions for VZERO-A detector obtained from 2 different sets of sub-events agrees within 2$\%$. This value is considered as a systematic uncertainty. Fig. \ref{fig1} right presents the event plane resolution of VZERO-A and VZERO-C detectors obtained with the 3 sub-event method as a function of the centrality of the collision. The resolution is maximal in semi-central events. 
Due to the limited statistics available in the data sample, the J/$\psi$ $v_{2}$ is extracted in wide centrality bins. The resolution in wide centrality bin is calculated from the resolution obtained in narrow centrality bins weighted by the number of reconstructed J/$\psi$ to take into account the fact that more J/$\psi$ are produced in central collisions than in peripheral collisions. 

\section{Results}

\begin{figure}[!htbp]
\begin{center}
\begin{minipage}[t]{.47\linewidth}
\hglue -0.5 true cm
 \includegraphics[width=1.\linewidth]{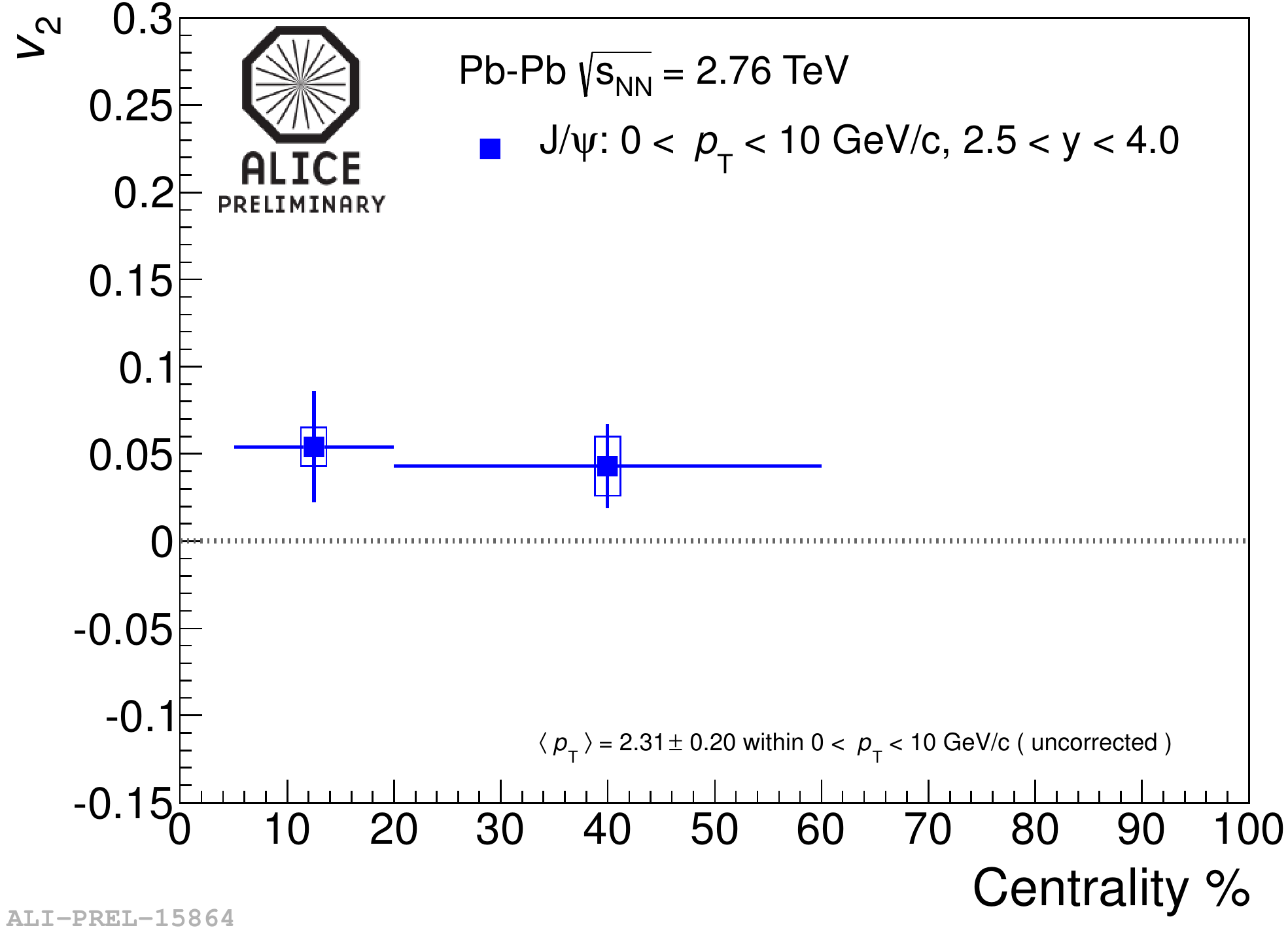}
\end{minipage}
\begin{minipage}[t]{.47\linewidth}
 \includegraphics[width=1.\linewidth]{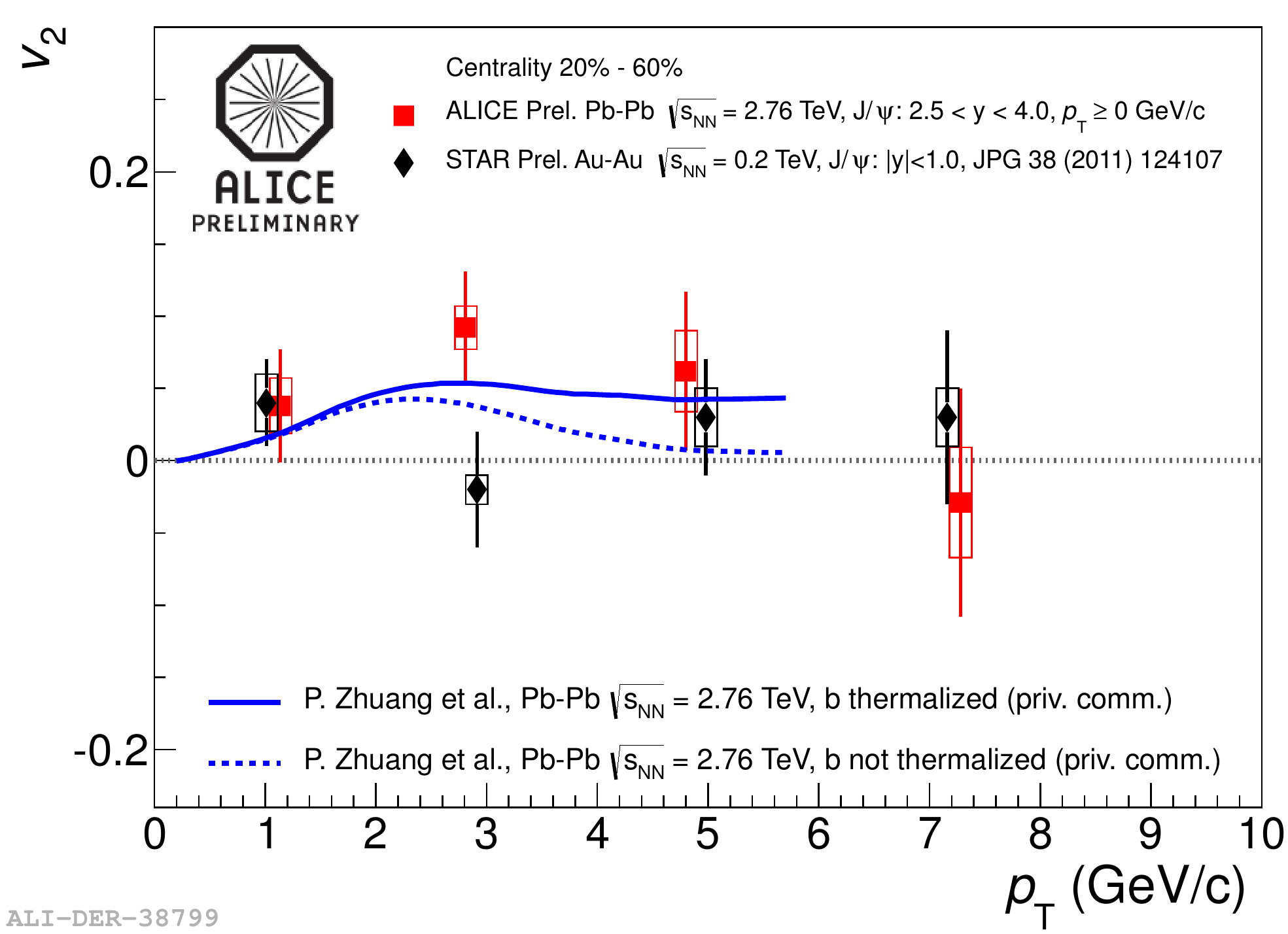}
\end{minipage}
\caption[fig1]{\label{fig2} \scriptsize{ Left: J/$\psi$ elliptic flow as a function of centrality in the rapidity range 2.5 $< y <$ 4 and in the $p_{\rm T}$ range 0 $< p_{\rm T} <$ 10 GeV/c. Boxes are the point to point uncorrelated systematic uncertainties. Crosses are the statistical uncertainties. Right: J/$\psi$ elliptic flow as a function of $p_{\rm T}$ in the centrality range 20-60$\%$ and in the rapidity range 2.5 $< y < 4$ (red squares). Crosses are the statistical uncertainties. Boxes are the point to point uncorrelated systematic uncertainties. A correlated systematic relative error of 2$\%$ is also considered on the event plane resolution. ALICE data are compared to STAR measurement performed in Au-Au collisions ($\sqrt{s_{NN}}$=200 GeV) and in the rapidity range  $|y| <$ 1 (black diamonds). ALICE data are compared with a transport model prediction (see text for details, colour online).}}
\end{center}
\vglue -1 true cm
\end{figure}

The integrated J/$\psi$ elliptic flow was extracted for two centrality bins (5-20$\%$ and 20-60$\%$ of the total hadronic cross section) in the $p_{\rm T}$ range 0 $< p_{\rm T} <$ 10 GeV/c and for the rapidity 2.5 $< y <$ 4 (see Fig. \ref{fig2} left). The two bins exhibit hints for a non-zero integrated flow. The mean $p_{\rm T}$ of the analyzed J/$\psi$ sample is 2.31 $\pm$ 0.20 GeV/c. The $p_{\rm T}$ dependence of the J/$\psi$ elliptic flow was studied up to 10 GeV/c for the centrality bin 20-60$\%$ (Fig. \ref{fig2} right). The measured $v_{2}$ is compatible with zero, except in the $p_{\rm T}$ range 2-4 GeV/c where its value amounts to $v_{2}$ = 0.092 $\pm$ 0.039 (stat) $\pm$ 0.015 (syst). Significance of non-zero flow in that $p_{\rm T}$ bin is 2.2$\sigma$. The main contribution to the systematic uncertainties comes from the signal extraction ($16\%$) in the $p_{\rm T}$ bin 2-4 GeV/c. The event plane method used in the analysis was validated by simulations. It was checked that if no flow is introduced at the simulation level, no J/$\psi$ elliptic flow is thus reconstructed in the muon spectrometer. Any deviation from zero J/$\psi$ elliptic flow observed in the simulation is considered as an additional source of systematic uncertainty in the analysis. In the $p_{\rm T}$ range 2 $< p_{\rm T} <$ 4 GeV/c, this contribution amounts to 0.1$\%$. This systematic uncertainty reaches 17$\%$ in the last $p_{\rm T}$ bin. A correlated systematic uncertainty of 2$\%$ coming from the event plane resolution determination is also considered. In the present analysis, the statistical uncertainty is the dominant source of error. ALICE data points (red squares) are also compared with STAR data points (black diamonds) obtained in Au-Au collisions ($\sqrt{s_{NN}}$ = 200 GeV) at mid-rapidity (Fig. \ref{fig2} right) and in the same centrality interval (20-60$\%$). A different behaviour is observed between STAR and ALICE in the $p_{\rm T}$ region 2 $< p_{\rm T} <$ 4 GeV/c where STAR measured a J/$\psi$ elliptic flow compatible with zero. ALICE data points are then compared with a transport model prediction \cite{THEO} (see Fig. \ref{fig2} right). This model assumes a production cross section of 0.38 mb for charm quark pairs and includes shadowing effects. It also assumes the thermalization of b quarks (blue line) or unthermalized b quarks (dashed blue line). The model predicts non zero J/$\psi$ elliptic flow in the intermediate $p_{\rm T}$ region. The magnitude of the predicted flow is comparable with the one measured. This model also describes the evolution of the J/$\psi$ nuclear modification factor versus centrality and transverse momentum measured at forward rapidity in ALICE \cite{Suire}.

\section{Conclusions}
We have presented the first measurement of inclusive J/$\psi$ elliptic flow in Pb-Pb collisions at $\sqrt{s_{NN}}$ = 2.76 TeV. Hints for non zero elliptic flow have been observed in the centrality-integrated bins 5-20$\%$ and 20-60$\%$. The $p_{\rm T}$ differential J/$\psi$ elliptic flow was extracted in the 20-60$\%$ centrality bin for four bins in transverse momentum, and compared to a preliminary result from STAR at $\sqrt{s_{NN}}$ = 200 GeV. Hints for non-zero elliptic flow, in the $p_{T}$ range 2 $< p_{\rm T} <$ 4 GeV/c has been observed with a significance of 2.2$\sigma$. The magnitude of the measured flow can be described by a parton transport model. Statistical uncertainties constitute the limitation of the present measurement.





\bibliographystyle{elsarticle-num}







\end{document}